\documentclass[]{article}

\usepackage{graphicx}
\usepackage{amsmath}
\usepackage{amssymb}
\usepackage{subfigure}
\usepackage{epsfig}
\usepackage{wrapfig}
\usepackage{color}
\usepackage{rotating}
\usepackage{dcolumn}
\usepackage{bm}
\usepackage{fullpage}

\usepackage[colorlinks=true, allcolors=blue]{hyperref}

\newcolumntype{d}[1]{D{.}{\cdot}{#1} }

\def\red{\textcolor{black}}

\def\##1{{\underline #1}}
\def\=#1{\underline{\underline{#1}}}
\def\+#1{\underline{\bf #1}}
\def\*#1{\breve{\bf #1}}

\def\.{\mbox{ \tiny{$^\bullet$} }}

\def\lec{\left\{}
\def\ric{\right\}}

\def\rr#1{\ref{#1}}

\def\eps{\varepsilon}
\def\epso{\eps_{\scriptscriptstyle 0}}
\def\etao{\eta_{\scriptscriptstyle 0}}
\def\ko{k_{\scriptscriptstyle 0}}
\def\lambdao{\lambda_{\scriptscriptstyle 0}}
\def\muo{\mu_{\scriptscriptstyle 0}}
\def\alphao{\alpha_{\scriptscriptstyle 0}}

\def\cpsi{\cos\psi}
\def\spsi{\sin\psi}

\def\stheta{\sin\theta}

\def\mn{^{(m,n)}}
\def\00{^{(0,0)}}
\def\m{^{(m)}}
\def\n{^{(n)}}

\def\ux{\hat{\#u}_x}
\def\uy{\hat{\#u}_y}
\def\uz{\hat{\#u}_z}

\def\sp{\#s}
\def\pinc{\#p_{\rm +}}
\def\pref{\#p_{\rm -}}

\def\Einc{\#E_{\rm inc}}
\def\Eref{\#E_{\rm ref}}
\def\Etrs{\#E_{\rm tr}}

\def\as{a_{\rm s}}
\def\ap{a_{\rm p}}
\def\rs{r_{\rm s}}
\def\rp{r_{\rm p}}
\def\ts{t_{\rm s}}
\def\tp{t_{\rm p}}

\def\rss{r_{\rm ss}}
\def\rsp{r_{\rm sp}}
\def\rps{r_{\rm ps}}
\def\rpp{r_{\rm pp}}
\def\tss{t_{\rm ss}}
\def\tsp{t_{\rm sp}}
\def\tps{t_{\rm ps}}
\def\tpp{t_{\rm pp}}

\def\Rss{R_{\rm ss}}
\def\Rsp{R_{\rm sp}}
\def\Rps{R_{\rm ps}}
\def\Rpp{R_{\rm pp}}
\def\Tss{T_{\rm ss}}
\def\Tsp{T_{\rm sp}}
\def\Tps{T_{\rm ps}}
\def\Tpp{T_{\rm pp}}

\def\As{A_{\rm s}}
\def\Ap{A_{\rm p}}

\def\deg{^\circ}

\def\calU{{\cal U}}

\def\mbbZ{\mathbb{Z}}

\def\kx{k_{\rm x}} 
\def\ky{k_{\rm y}} 
\def\kxy{k_{\rm xy}}

\def\alphamet{\bar{A}^{\rm met}}
\def\alphasc{\bar{A}^{\rm sc}}
\def\alphatot{\bar{A}^{\rm tot}}
\def\La{L_{\rm a}}	
\def\Ld{L_{\rm d}}	
\def\Lg{L_{\rm g}}	
\def\Lm{L_{\rm m}}
\def\Ls{L_{\rm s}}	
\def\Lt{L_{\rm t}}
\def\Lw{L_{\rm w}}
\def\Lx{L_{\rm x}}
\def\Ly{L_{\rm y}}
\def\zetax{\zeta_{\rm x}}
\def\zetay{\zeta_{\rm y}}

\def\ppol{{\textit{p}}$-$\rm polarized}

\def\Pin{$p$-$i$-$n$}
\def\pin{\Pin~}
\def\Assc{\bar{A}^{\rm sc}_{\rm s}}
\def\Apsc{\bar{A}^{\rm sc}_{\rm p}}

\def\Aassc{\bar{A}^{\rm sc1}_{\rm s}}
\def\Aapsc{\bar{A}^{\rm sc1}_{\rm p}}
\def\Aaassc{\bar{A}^{\rm sc2}_{\rm s}}
\def\Aaapsc{\bar{A}^{\rm sc2}_{\rm p}}
\def\Aaaassc{\bar{A}^{\rm sc3}_{\rm s}}
\def\Aaaapsc{\bar{A}^{\rm sc3}_{\rm p}}

\def\epsw{\eps_{\rm w}}
\def\epsd{\eps_{\rm d}}
\def\epsm{\eps_{\rm m}}
\def\epsg{\eps_{\rm g}}

\begin{document}
\begin{center}
\textbf{On optical-absorption peaks in a nonhomogeneous thin-film solar cell with a two-dimensional periodically corrugated metallic backreflector  }\\

\textit{Faiz Ahmad,$^a$ 
	Tom H. Anderson,$^b$
	Benjamin J. Civiletti,$^b$
	Peter B. Monk,$^b$ and
	Akhlesh Lakhtakia$^{a,*}$}\\

{$^a$Pennsylvania State University, Department of Engineering Science and Mechanics, NanoMM--Nanoengineered Metamaterials Group,   University Park, PA 16802, USA\\
	$^b$University of Delaware, Department of Mathematical Sciences,
	501 Ewing Hall, \\ Newark, DE 19716, USA\\
	$^*${akhlesh@psu.edu}}
	\end{center}

	\begin{abstract}
		The rigorous coupled wave approach (RCWA) was implemented to investigate   
		optical absorption in a triple-\Pin-junction amorphous-silicon solar cell 
		with 
		a 2D metallic periodically corrugated backreflector (PCBR). 
		Both total and useful 
		absorptances 
		were computed against the free-space wavelength $\lambdao$
		for both $s$- and $p$-polarized polarization states. 
		The useful absorptance 
		in each of the three \Pin~junctions was also computed for 
		normal as well as oblique incidence.  
		Furthermore, two canonical boundary-value problems
		were solved for the prediction of guided-wave modes (GWMs):
		surface-plasmon-polariton  waves and waveguide modes.
		Use of the doubly periodic PCBR enhanced both useful and total absorptances
		in comparison to a planar backreflector.
		The predicted GWMs were correlated 
		with the peaks of the total and useful absorptances.
		The excitation of GWMs was mostly confined to $\lambdao<700$~nm
		and enhanced absorption. As
		excitation of certain GWMs could be correlated with the total absorptance 
		but not with the useful absorptance,  the useful absorptance should be 
		studied while devising light-trapping strategies. 
	\end{abstract}

	\section{Introduction}
	
	Amorphous silicon (a-Si) thin-film solar cells provide a viable option 
	to the 1$^{\rm st}$-generation crystalline-silicon (c-Si) solar cells \cite{Singh2009}, 
	due to their ease of manufacturing and low cost.  But the typical efficiency of 
	a-Si thin-film
	solar cells is not as high as of c-Si solar cells, due to the  high electron-hole 
	recombination rate and low charge-carrier diffusion lengths in a-Si \cite{1, 2}.
	Consequently, light-trapping techniques are necessary to enhance the  efficiency 
	of a-Si thin-film solar cells. Several light-trapping strategies have been studied both 
	experimentally and theoretically\cite{3,MPL2017}. Anti-reflection coatings \cite{Kavakli2002,4,5}, 
	textured front faces \cite{Southwell1991, Sahoo2009}, metallic
	periodically corrugated  backreflectors (PCBRs) \cite{Sheng,Heine,7}, 
	particle plasmonics\cite{ZJOG}, surface plasmonics \cite{Anderson1982,Anderson1983,Deceglie} 
	and multiplasmonics \cite{6,SBLFMM,LBSYLMM}, and
	waveguide-mode excitation \cite{9,10,11} are attractive for trapping light in  solar cells.

	Of particular interest is the enhancement of the optical electric field through the excitation of two types of guided-wave modes (GWMs):
	surface-plasmon-polariton (SPP) waves and wave\-guide modes (WGMs). 
	The periodically corrugated interface of a metal and a semiconductor that is periodically nonhomogeneous in the thickness direction (identified by the $z$ axis in Sec.~\rr{sec:theory})  can guide multiple SPP
	waves at the same frequency\cite{6,ESW2013}. Any open-face waveguide with an air/semiconductor/metal architecture can guide WGMs \cite{10,11,14}. Therefore, the incorporation of nonhomogeneity along the thickness direction in the semiconductor layers of a solar cell with a PCBR can enhance photonic absorption
	\cite {6,9,Anderson2016}. That enhancement would increase the generation rate of electron-hole pairs \cite{Anderson2017,Anderson2016}.
	
	Much of the theoretical and experimental research done on thin-film solar cells
	with metallic PCBRs is confined to devices with a homogeneous semiconductor layer and a metallic backreflector with one-dimensionally (1D) periodic corrugation.
	An experimental report of broadband excitation of multiple SPP waves in a device comprising a 1D photonic crystal (PC) atop a 1D PCBR \cite{Hallnano} confirmed theoretical predictions \cite{Faryad-Hall} and spurred research on solar cells
	containing piecewise nonhomogeneous semiconductor layers and 1D PCBRs
	\cite{6,Anderson2016,Anderson2017,FLML}.
	In a recent study, experimental excitation of multiple SPP waves and WGMs were reported in  a device comprising a 1D~PC atop a 2D~PCBR
	\cite{9}. Appropriately
	designed 2D~PCBRs were found to be better for the excitation of GWMs
	than 1D~PCBRs, after
	the broadband excitation of GWMs predicted by solving two canonical boundary-value problems was correlated with the experimentally measured absorption spectrums.
	
	In solar-cell research, often the excitation of GWMs is correlated with the total absorptance $\alphatot$ of the device \cite{6,11}, which however is not a good measure  of useful photonic absorption in a solar cell,  as photons absorbed in the metallic portions of a solar cell are not available for conversion into electric current. 
	Therefore, the chief objective for the work reported in this paper was to  determine the spectrums of both the total absorptance $\alphatot$ and the useful absorptance $\alphasc$
	\cite{SFSML} in a tandem solar cell with a 2D~PCBR exposed to either normally or obliquely incident linearly polarized light. The solar cell was taken to comprise three \pin solar cells
	made of a-Si alloys \cite{16} that can be fabricated using plasma-enhanced chemical-vapor deposition over planar and patterned substrates. A top layer of aluminum-doped zinc oxide (AZO) was incorporated to provide a transparent electrode. Also, an AZO layer was taken to be sandwiched between the 2D~PCBR and the stack of nine semiconductor layers in order to avoid the deterioration of the electrical properties of the a-Si alloy closest to the metal \cite{22}, which was chosen to be silver \cite{17}. The total absorptance and the useful absorptance calculated using the rigorous coupled-wave approach (RCWA)
	\cite{GG,Onishi,ESW2013}
	were correlated against the predicted excitations of GWMs.

	The plan of this paper is as follows. Section~\rr{sec:theory} is divided into four parts. Section~\rr{sec:reftransabs} presents the boundary-value
	problem that can be solved to determine the optical \red{electromagnetic} fields everywhere in a device comprising a stratified, isotropic dielectric material atop a 2D~PCBR, when the
	device is illuminated by a plane wave. The formulations for useful and total absorptances are discussed in Sec.~\rr{absorpt}. Section~\rr{canprob} provides  brief descriptions of the underlying canonical problems to predict the excitation of SPP waves and WGMs. Excitation of GWMs is discussed in Sec.~\rr{GWMexcit}. Section~\rr{nrd} is divided into two parts. The wavenumbers of the predicted GWMs are presented in Sec.~\rr{predGWM}. Correlations of the absorptances with the predicted GWMs are discussed in Sec.~\rr{abscorrel}.
	The paper concludes with some remarks in Sec.~\rr{conc}.
	
	An $\exp{(-i \omega t)}$    dependence on time $t$ is implicit, with $\omega$ denoting the angular frequency and $i=\sqrt{-1}$. The free-space wavenumber, the free-space wavelength, and the intrinsic impedance of free space are denoted by $\ko=\omega \sqrt{\muo\epso}$, $\lambdao=2\pi/\ko$, and $\etao=\sqrt{\muo/\epso}$, respectively, with $\muo$ being the permeability and $\epso$ the permittivity of  free space. Vectors are underlined; the Cartesian unit vectors are identified as $\ux$, $\uy$, and $\uz$; and column vectors as well as matrixes are in boldface.


	\section{Theory in Brief}\label{sec:theory}
	
	\subsection{Boundary-value problem for tandem solar cell} \label{sec:reftransabs}
	
\begin{figure} 
		\centering
		\includegraphics[scale=0.4]{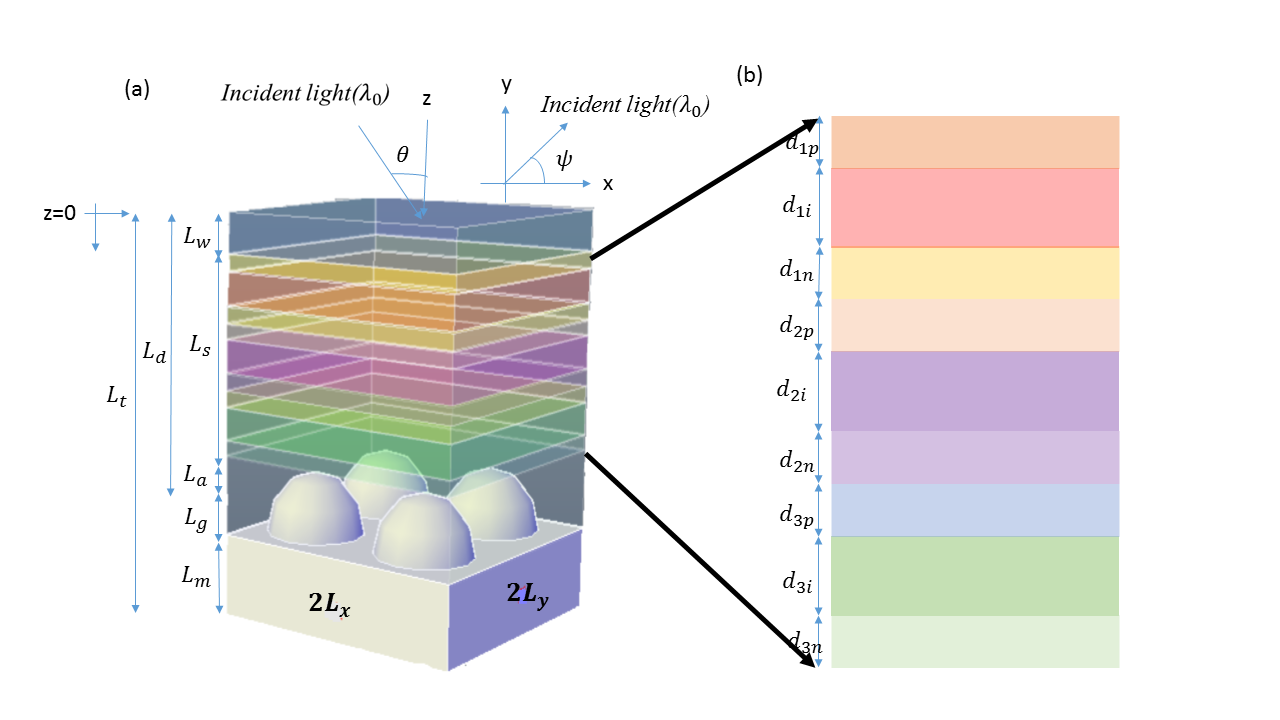}
		\caption{ (a) Schematic of the tandem solar cell comprising three \pin junctions of a-Si alloys on a 2D~PCBR. 
			The wavevector of the incident	plane wave is inclined at angle $\theta$ with respect to the $z$
			axis and angle  $\psi$ with respect to the $x$ axis in the $xy$ plane.   (b) 
			Nine semiconductors layers of the three \pin junctions.
			\label{figure1}}
	\end{figure}

	Let us consider the boundary-value problem shown in Fig.~\ref{figure1} for a tandem solar cell containing
	three \pin junctions. The solar cell occupies the region ${\cal X}:\left\{(x,y,z)
	\vert -\infty<x<\infty, -\infty<\right.$
	$\left.
	y<\infty, 0<z<\Lt\right\}$, with the half spaces $z<0$ and $z>\Lt$ occupied by air.
	The reference unit cell is identified as 
	${\cal R}:\left\{(x,y,z)
	\vert -\Lx/2<x<\Lx/2,\right.$ $\left. -\Ly/2<y<\Ly/2,  0<z<\Lt\right\}$, 
	the backreflector being  periodic along both the $x$ and $y$ axes.
	
	The region $0<z<\Ld=\Lw+\Ls+\La$ is occupied by a cascade
	of homogeneous layers and is compactly characterized 
	by the   permittivity $\epsd(z,\lambdao)$, which is a piecewise constant function of $z$. The top layer  $0<z<\Lw$ and
	the bottom layer $\Lw+\Ls<z<\Ld$ are
	made of AZO with  permittivity $\epsw(\lambdao)$. 
	The semiconductor layers in the region $\Lw<z<\Lw+\Ls$
	are identified in Fig.~\ref{figure1}(b).  
	The region $\Ld+\Lg < z < \Ld+\Lg+\Lm$ is occupied by a metal with  permittivity $\epsm(\lambdao)$.

	The region $\Ld < z < \Ld  +\Lg$,
	henceforth termed the grating region, contains a periodically undulating surface 
	with period $\Lx$ along the $x$ axis and   period $\Ly$ along the $y$ axis.  
	In the grating region, $\cal X$ possesses    rectangular  symmetry in the $xy$ plane.
	The  permittivity $\epsg(x,y,z,\lambdao)$ in the grating region  can be stated as
	\begin{eqnarray}
	\nonumber
	&&\epsg(x,y,z,\lambdao) =
	\epsm(\lambdao)+[\epsw(\lambdao)- \epsm(\lambdao)] \,\calU[z-g_1(x)]\calU[z-g_2(y)]\,, 
	\\[4pt]
	&&\qquad\qquad \vert{x}\vert<\zetax \Lx/2\,,
	\quad \vert{y}\vert<\zetay\Ly/2\,,\quad z\in(\Ld,\Ld+\Lg)\,,
	\label{eq:perm}
	\end{eqnarray}
	where the unit step function
	\begin{equation}
	\calU(\sigma)=
	\lec\begin{array}{ll}
	0\,, &\sigma < 0,
	\\
	1\,, &\sigma \geq0, 
	\end{array}\right.
	\end{equation}
	and $\zetax \in[0,1]$ as well as $\zetay \in [0,1] $ are the duty cycles.
	We chose the grating-shape functions
	\begin{equation}  
	\displaystyle{
		g_1(x)= \left\{
		\begin{array}{ll}
		\Ld+\Lg[1-\cos( 2\pi \frac{\pi x}{\zetax \Lx})] \, , &x\in [-\frac{\zetax \Lx}{2},\frac{\zetax \Lx}{2} ] ,\\[5pt]
		\Ld+\Lg \, , &x\notin [-\frac{\zetax \Lx}{2},\frac{\zetax \Lx}{2} ] ,
		\end{array}  \right.
	}
	\end{equation}
	and
	\begin{equation}   
	\displaystyle{ 
		g_2(y)=
		\left\{
		\begin{array}{ll}
		\Ld+\Lg[1-\cos( 2\pi \frac{\pi y}{\zetay \Ly})] \, , &y\in [-\frac{\zetay \Ly}{2},\frac{\zetay \zetay}{2} ], \\[5pt]
		\Ld+\Lg \, , &y\notin [-\frac{\zetay \Ly}{2},\frac{\zetay \Ly}{2} ],  \end{array}  
		\right.
	}
	\end{equation} 
	to represent hillocks for all data reported in this paper. The grating-shape functions chosen here are only for illustration,  many other choices fit for experimental study being also available \cite{7}.
	
	Suppose that an arbitrarily polarized plane wave, propagating in the half space $z<0$ at an angle $\theta\in[0\deg,90\deg)$ 
	with respect to the $z$ axis and  an angle $\psi\in[0^\circ,360^\circ)$ with respect
	to the $x$ axis in the $xy$ plane, is incident on the plane $z = 0$. 
	The electric field phasor of this plane wave can be stated as
	\begin{equation}
	\Einc(\#r)=\left(\bar{a}_{\rm s}\, \sp\00+\bar{a}_{\rm p}\, \pinc\00\right)
	\exp\left[i\left(\#\kappa\00+\alpha_0\00 \uz \right)\. \#r\right]\,,
	\end{equation}
	where $\bar{a}_{\rm s}$ and $\bar{a}_{\rm p}$ are the known coefficients
	of $s$- and $p$-polarized components, respectively.
	Here and hereafter, the following quantities are used:
	\begin{equation} \left. \displaystyle{\begin{array}{l}
		\#\kappa\mn=\kx\m \ux+\ky\n \uy
		\\[4pt]
		\kx\m =\ko\stheta\cpsi +m(2\pi/\Lx)
		\\[4pt]
		\ky\n =\ko\stheta\spsi +n(2\pi/\Ly)
		\\[4pt]
		\kxy\mn=+\sqrt{\#\kappa\mn\.\#\kappa\mn}\\[5pt]
		\alphao\mn=+\sqrt{\ko^2-\#\kappa\mn\.\#\kappa\mn}\\[5pt]
		\displaystyle{\sp\mn=-\frac{\ky\n}{\kxy\mn}\ux+\frac{\kx\m}{\kxy\mn} \uy}\\[10pt]
		\displaystyle{\pinc\mn=-\left(\frac{\kx\m}{\kxy\mn}\ux+\frac{\ky\n}{\kxy\mn}\uy\right)\frac{\alphao\mn}{\ko}+\frac{\kxy\mn}{\ko}\uz}\\[10pt]
		\displaystyle{\pref\mn=\left(\frac{\kx\m}{\kxy\mn}\ux+\frac{\ky\n}{\kxy\mn}\uy\right)\frac{\alphao\mn}{\ko}+\frac{\kxy\mn}{\ko}\uz}
		\end{array}}    \ric\,,\quad {m\in\mathbb{Z}}\, ,\quad {n\in\mathbb{Z}}\,.
	\end{equation}

	As a result of
	the metallic PCBR being doubly  periodic,
	the $x$- and $y$-dependences of the electric and magnetic field phasors  are represented everywhere as an infinite series of Floquet harmonics as \cite{GG,Onishi,ESW2013}
	\begin{equation}
	\left.\begin{array}{l}
	\displaystyle{
		{\#E}(x,y,z) = \sum_{m\in\mathbb{Z}}\sum_{n\in\mathbb{Z}}
		{\#e}^{(m,n)}(z)\,
		\exp\left(i\#\kappa\mn\. \#r\right) 
	}
	\\[6pt]
	\displaystyle{
		{\#H}(x,y,z) = \sum_{m\in\mathbb{Z}}\sum_{n\in\mathbb{Z}}
		{\#h}^{(m,n)}(z)\,
		\exp\left(i\#\kappa\mn\. \#r\right)
	}\end{array}
	\right\}\,,
	\label{eq6}
	\end{equation}
	where 
	\begin{equation}
	\left.\begin{array}{l}
	\displaystyle{{\#e}^{(m,n)} (z)= {e}_x^{(m,n)}(z)\ux+{e}_y^{(m,n)}(z)\uy+{e}_z^{(m,n)}(z)\uz}
	\\[6pt]
	\displaystyle{
		{\#h}^{(m,n)}(z)= {h}_x^{(m,n)}(z)\ux+{h}_y^{(m,n)}(z)\uy+{h}_z^{(m,n)}(z)\uz}
	\end{array}
	\right\}\,
	\label{eq7}
	\end{equation}
	are  expansion coefficients.
	Accordingly, the incident and the reflected electric field phasors are represented as
	\begin{equation}
	\left.\begin{array}{l}
	\Einc(x,y,z)=\displaystyle{\sum_{m\in \mbbZ}\sum_{n\in \mbbZ}}
	\left\{
	\left({ \as\mn\#s\mn}+\ap\mn\pinc\mn \right) 
	\right.
	\\[8pt]
	\left.\qquad\times\exp\left[i\left(\#\kappa\mn+\alpha_0\mn \uz \right)\.\#r\right]
	\right\}
	\\[8pt]
	\Eref(x,y,z)=\displaystyle{\sum_{m\in \mbbZ}\sum_{n\in \mbbZ}}
	\left\{
	\left({ \rs\mn\#s\mn}+\rp\mn\pref\mn \right)
	\right.
	\\[8pt]
	\left.\qquad\times\exp\left[i\left(\#\kappa\mn-\alpha_0\mn \uz \right)\.\#r\right]
	\right\}
	\end{array}\right\}
	\,, \quad z< 0\,,
	\label{Eincref}
	\end{equation}
	and the transmitted electric field phasor as
	\begin{eqnarray}
	\nonumber
	\Etrs(x,y,z)&=&\displaystyle{\sum_{m\in \mbbZ}\sum_{n\in \mbbZ}}
	\left\{
	\left({ \ts\mn\#s\mn}+\tp\mn\pinc\mn \right)\right.\\
	&&\qquad \left.\times\exp\left[i\left(\#\kappa\mn+\alpha_0\mn \uz \right)\.
	\left(\#r-\Lt\uz\right)\right]\right\}\,, \quad z>\Lt\,,
	\label{Etr}
	\end{eqnarray}
	\red{where} the coefficients
	$\as\mn=\bar{a}_{\rm s}\delta_{m0}\delta_{n0}$ and
	$\ap\mn=\bar{a}_{\rm p}\delta_{m0}\delta_{n0}$
	in Eq.~(\ref{Eincref})$_1$ are known with $\delta_{mm^\prime}$ denoting
	the Kronecker delta, \red{but}
	the coefficients $\rs\mn$, $\rp\mn$, $\ts\mn$, and $\tp\mn$ in
	Eq.~(\ref{Eincref})$_2$ and Eq.~(\ref{Etr}) have to be determined.
	Finally, the  permittivity $\eps(x,y,z)$ everywhere is represented by the Fourier series
	\begin{equation}
	\displaystyle{
		\eps(x,y,z) = \displaystyle{\sum_{m\in\mathbb{Z}}\sum_{n\in\mathbb{Z}}}
		{\eps}^{(m,n)}(z)\exp\left[i\left(\#\kappa\mn-\#\kappa\00\right)\. \#r\right]
	}\,,
	\label{eq8}
	\end{equation}
	where  ${\eps}^{(m,n)}(z)$ are Fourier coefficients.

	Computational tractability requires the  expansions in Eqs.~(\ref{eq6})--(\ref{eq8}) to be truncated to include only  
	$m \in \left\{-M_t,...,M_t\right\}$ and $n \in \left\{-N_t,...,N_t\right\}$, 
	with $M_t\geq0$ and $N_t\geq0$. Furthermore, a superindex 
	\begin{equation}
	\tau=m (2 N_t+1)+n \,,\quad
	m\in\left[-M_t,M_t\right]\,,\quad
	n\in\left[-N_t,N_t\right]\,,
	\end{equation}
	is defined for convenience. Then,
	$\tau\in\left[-\tau_t,\tau_t\right]$, where $\tau_t=2M_tN_t+M_t+N_t$. Also, both
	the mapping from $(m,n)$ to $\tau$ 
	and the inverse mapping from $\tau$  to $(m,n)$ are injective  \cite{Kreyszig}.
	Thereafter, column vectors
	\begin{equation}
	\left.\begin{array}{l}
	\*e_\sigma(z)=\left[ 
	e_\sigma^{(-\tau_t)}(z),e_\sigma^{(-\tau_t+1)}(z),...,e_\sigma^{(\tau_t-1)}(z),e_\sigma^{(\tau_t)}(z)\right]^T
	\\[5pt]
	\*h_\sigma(z)=\left[ 
	h_\sigma^{(-\tau_t)}(z),h_\sigma^{(-\tau_t+1)}(z),...,h_\sigma^{(\tau_t-1)}(z),h_\sigma^{(\tau_t)}(z)\right]^T
	\end{array}\right\}\,
	,\quad\sigma\in\{x,y,z\},
	\end{equation}
	of length $2\tau_t+1$
	are set up, the superscript $T$ denoting the transpose.  The Toeplitz matrix \cite{HBM}
	\begin{equation}
	\red{\mathbf{\breve{\boldsymbol{\eps}}}(z)}=
	\begin{bmatrix}
	\breve{\eps}^{(-\tau_{t},-\tau_{t})}(z)       
	& \breve{\eps}^{(-\tau_{t},-\tau_{t}+1)}(z)  & \cdots  
	& \breve{\eps}^{(-\tau_{t},\tau_{t}-1)}(z) 
	& \breve{\eps}^{(-\tau_{t},\tau_{t})}(z)
	\\
	\breve{\eps}^{(-\tau_{t}+1,-\tau_{t})}(z)        
	& \breve{\eps}^{(-\tau_{t}+1,-\tau_{t}+1)}(z)  & \cdots  
	& \breve{\eps}^{(-\tau_{t}+1,\tau_{t}-1)}(z)   
	& \breve{\eps}^{(-\tau_{t}+1,\tau_{t})}(z)
	\\
	\cdots      & \cdots  & \cdots  & \cdots & \cdots 
	\\
	\breve{\eps}^{(\tau_{t}-1,-\tau_{t})}(z)        
	& \breve{\eps}^{(\tau_{t}-1,-\tau_{t}+1)}(z) & \cdots  
	& \breve{\eps}^{(\tau_{t}-1,\tau_{t}-1)}(z)  
	& \breve{\eps}^{(\tau_{t}-1,\tau_{t})}(z)
	\\
	\breve{\eps}^{(\tau_{t},-\tau_{t})}(z)        
	& \breve{\eps}^{(\tau_{t},-\tau_{t}+1)}(z) & \cdots  
	& \breve{\eps}^{(\tau_{t},\tau_{t}-1)}(z)  
	& \breve{\eps}^{(\tau_{t},\tau_{t})}(z) 
	\end{bmatrix}.
	\end{equation}
	contains the
	Fourier coefficients appearing in Eq.~(\ref{eq7}) with
	\red{$\breve{\eps}^{(\tau,\tau^\prime)}(z)={\eps}^{(m-m^\prime,n-n^\prime)}(z)$}.
	Finally, the $(2\tau_t+1)\times(2\tau_t+1)$ Fourier-wavenumber matrixes 
	\begin{equation}
	\left.\begin{array}{l}
	\displaystyle{
		\*K_{\rm x} = {\rm diag}\left[ \breve{k}_{\rm x}^{(-\tau_t)},\, \breve{k}_{\rm x}^{(-\tau_t+1)},\,...,
		\breve{k}_{\rm x}^{(\tau_t-1)},\,\breve{k}_{\rm x}^{(\tau_t)}\right]
	}
	\\[6pt]
	\displaystyle{
		\*K_{\rm y} = {\rm diag}\left[ \breve{k}_{\rm y}^{(-\tau_t)},\, \breve{k}_{\rm y}^{(-\tau_t+1)},\,...,
		\breve{k}_{\rm y}^{(\tau_t-1)},\,\breve{k}_{\rm y}^{(\tau_t)}\right]
	}
	\end{array}\right\}
	\end{equation}
	are set up with 
	$\breve{k}_{\rm x}^{(\tau)} = k_{\rm x}^{(m)}$ and
	$\breve{k}_{\rm y}^{(\tau)} = k_{\rm y}^{(n)}$.

	The frequency-domain Maxwell curl \red{postulates  yield} the
	matrix ordinary differential equation \cite{ESW2013}
	\begin{equation}
	\frac{d}{dz} \*f(z) =i\*P(z)\.\*f(z), 
	\label{eq:ode}
	\end{equation}  
	where the $4(2\tau_t+1)$-column vector  
	\begin{equation}
	\*f(z)=\left[  \begin{array}{c}
	\*e_x(z)\\[3pt]
	\*e_y(z)\\[3pt]
	\*h_x(z)\\[3pt]
	\*h_y(z)
	\end{array}
	\right]
	\end{equation}
	and the $4(2\tau_t+1)\times4(2\tau_t+1)$ matrix  
	\begin{equation}
	\nonumber
	\*P(z)=\omega\left[
	\begin{array}{cccc}
	\*0&\*0&\*0&\muo\*I\\[3pt]
	\*0&\*0&- \muo\*I&\*0\\[3pt]
	\*0& - \*{\boldsymbol \eps}(z) &\*0&\*0\\[3pt]
	\*{\boldsymbol \eps}(z)&\*0&\*0&\*0
	\end{array}\right] 
	\end{equation}
	\begin{equation}
	+\frac{1}{\omega}\left[
	\begin{array}{cccc}
	\*0&\*0
	&\*K_x\.\left[\*{\boldsymbol \eps}(z)\right]^{-1}\.\*K_y
	&-  \*K_x\.\left[\*{\boldsymbol \eps}(z)\right]^{-1}\.\*K_x\\[3pt]
	\*0&\*0
	&\*K_y\.\left[\*{\boldsymbol \eps}(z)\right]^{-1}\.\*K_y
	& - \*K_y\.\left[\*{\boldsymbol \eps}(z)\right]^{-1}\.\*K_x
	\\[3pt]
	-\muo^{-1}\*K_x\.\*K_y
	&\muo^{-1}\*K_x\.\*K_x  
	&\*0&\*0\\[3pt]
	-\muo^{-1}\*K_y\.\*K_y
	& \muo^{-1}\*K_y\.\*K_x  
	&\*0&\*0
	\end{array}\right]
	\label{eq:eigenvalues}
	\end{equation}contains  $\*0$ as the $(2\tau_t + 1)\times(2\tau_t+1)$ null matrix and
	$\*I$ as the $(2\tau_t + 1)\times(2\tau_t+1)$ identity matrix. 
	
	In order to solve Eq.~(\ref{eq:ode}), the region $\cal R$ is partitioned into a sufficiently large number of  thin slices along the $z$ direction \cite{ESW2013}. Each  slice is taken to be homogeneous along the $z$ axis but it is either homogeneous or periodically nonhomogeneous along the $x$ and $y$ axes; thus, $\*P(z)$ is assumed to be uniform in each slice.  
	Boundary conditions are enforced on the planes $z=0$ and $z=\Lt$ to match the fields to the incident, reflected, and \red{transmitted  waves,} as appropriate. A stable numerical marching algorithm is then used to determine the Fourier coefficients of the electric and magnetic field phasors in each slice\cite{ESW2013}.
	Finally, the $z$ components of the electric and magnetic field phasors in the device can be obtained through
	$\*e_z(z)= -\left[\omega{\boldsymbol\eps}(z)\right]^{-1}\. 
	[\*K_x\.\*h_y(z)-\*K_y\.\*h_x(z)]$ and
	$\*h_z(z)= (\omega\muo)^{-1}[\*K_x\.\*e_y(z)-\*K_y\.\*e_x(z)]$.
	Thus, the electric field phasor can be determined everywhere. 
	The entire procedure was implemented on the 
	Mathematica$^{\textregistered}$~platform.
	
	\subsection{Total and useful absorptances}\label{absorpt}
	At any location inside the device, the absorption rate of the monochromatic optical energy per unit volume is given by 
	\begin{equation}
	Q(x,y,z)=\frac{1}{2} \omega\,{\rm Im}\{\eps(x,y,z)\}\left\vert\#E(x,y,z)\right\vert^2.
	\end{equation} 
	The useful absorptance   \cite{Ahmad-SPIE2017}
	\begin{equation}
	\alphasc=\frac{2\etao}{\Lx \Ly\left(
		\left\vert{\bar{a}_{\rm s}}\right\vert^2 +\left\vert{\bar{a}_{\rm p}}\right\vert^2\right)
		\cos\theta} \iiint_{{\cal R}_{\rm sc}} Q(x,y,z)\,dx\,dy\,dz\,
	\end{equation} 
	is calculated by integrating $Q(x,y,z)$ over the region
	${\cal R}_{\rm sc}\subset {\cal R}$ occupied
	by the semiconductor  layers. Likewise, absorptance in the metal is given by  
	\begin{equation}
	\alphamet=\frac{2\etao}{\Lx \Ly\left(
		\left\vert{\bar{a}_{\rm s}}\right\vert^2 +\left\vert{\bar{a}_{\rm p}}\right\vert^2\right)
		\cos\theta} \iiint_{{\cal R}_{\rm met}} Q(x,y,z)\,dx\,dy\,dz\,,
	\end{equation} 
	where  ${\cal R}_{\rm met}\subset {\cal R}$ is the region occupied
	by the metal. The total absorptance is then the sum
	\begin{equation}
	\alphatot=\alphasc+\alphamet\,,
	\end{equation}
	if $\epsw$ is purely real.
	
	Four reflection and four transmission coefficients of order $(m,n)$ are defined as the elements in the  2$\times$2  matrices appearing
	in the following relations \cite{ESW2013}:
	\begin{equation}
	\begin{bmatrix}
	\rs\mn\\[4pt] \rp\mn \end{bmatrix}= \begin{bmatrix}
	\rss\mn &\rsp\mn \\[4pt] \rps\mn & \rpp\mn \end{bmatrix}\.\begin{bmatrix}
	\bar{a}_{\rm s}\\[4pt] \bar{a}_{\rm p} \end{bmatrix}\,,\qquad
	\begin{bmatrix}
	\ts \mn\\[4pt] \tp\mn \end{bmatrix}=\begin{bmatrix}
	\tss\mn &\tsp\mn \\[4pt]\tps\mn & \tpp\mn \end{bmatrix}\.\begin{bmatrix}
	\bar{a}_{\rm s}\\[4pt] \bar{a}_{\rm p} \end{bmatrix}\,.
	\end{equation}
	Coefficients of order $(0,0)$ are classified as specular, whereas coefficients of all other orders are nonspecular. Four
	reflectances and four linear transmittances of order $(m,n)$ are defined as 
	\begin{equation}
	\displaystyle{
		\Rsp\mn= \frac{{\rm Re}\left[\alpha_0\mn\right] }{\alpha_0\00}
		\left\vert\rsp\mn\right\vert^2
	}
	\in[0,1]\,,
	\end{equation}
	etc., and two  absorptances as
	\begin{equation}
	\left.\begin{array}{l}
	\displaystyle{
		\As= 1-\sum_{m=-M_t}^{m=M_t}\sum_{n=-N_t}^{n=N_t}
		\left(\Rss\mn+\Rps\mn+\Tss\mn+\Tps\mn\right)  \in[0,1]
	}
	\\[8pt]
	\displaystyle{
		\Ap= 1-\sum_{m=-M_t}^{m=M_t}\sum_{n=-N_t}^{n=N_t}
		\left(\Rpp\mn+\Rsp\mn+\Tpp\mn+\Tsp\mn\right)  \in[0,1]
	}
	\end{array}
	\right\}\,.
	\end{equation}
	These are total absorptances in that they contain the contributions of the semiconductors and the metal in the solar cell.
	Whereas $\alphatot$, $\alphasc$, and $\alphamet$ are defined for incident light of arbitrary polarization state,
	$\As$ is defined for incident $s$-polarized light and $\Ap$ for incident $p$-polarized light. All  absorptances presented in Sec.~\ref{nrd} were calculated for a solar cell comprising just one triple \pin~junction, as shown in Fig.~\ref{figure1}.

	\subsection{Canonical boundary-value problems}\label{canprob}
	Two separate canonical boundary-value problems were solved to correlate
	peaks in the spectrums of various absorptances with the 
	excitation of SPP waves and WGMs. Details on both canonical
		problems are available elsewhere \cite{Ahmad-SPIE2017} for the interested
		reader, but we note the following salient features of both canonical problems.
	
	\subsubsection{SPP waves} The complex-valued  wavenumbers $q\ne0$
	of SPP waves for a specific value of $\lambdao$ were obtained by solving a canonical boundary-value problem \cite{ESW2013,Faryad-Hall}, with the assumptions that the backreflector metal occupies the half space $z<0$, a periodically semi-infinite cascade of three \pin junctions occupies
	the half space $z>0$, and there are no AZO layers.
	
	\subsubsection{Waveguide modes} An open-faced waveguide is formed by the three  \pin junctions interposed between two half spaces, one occupied by air and the other by the backreflector metal of thickness considerably exceeding the skin depth \cite{Iskander}.  For a specific value of $\lambdao$,
	this waveguide can support the propagation of multiple WGMs (with wavenumbers
	$q\ne0$)  which can play 
	significant light-trapping roles \cite{9,10,11}. We ignored the AZO layers for this canonical problem as well.\\

	\subsection{Excitation of SPP waves and WGMs}\label {GWMexcit}
	Planewave illumination will excite a GWM of wavenumber $q$
	as a Floquet harmonic of order $(m,n)$, provided that\cite{ESW2013}
	\begin{equation}
	\pm {\rm Re}\left[q\right]\simeq \kxy\mn\,.
	\label{predict0}
	\end{equation}
	When $\Lx=\Ly=L$, the right side of Eq.~(\ref{predict0}) simplifies to yield
	\begin{equation}
	\pm {\rm Re}\left[q/\ko\right]\simeq \left\{\left[\sin\theta+(m\cos\psi+n\sin\psi)(\lambdao/L)\right]^2+\left[(m\sin\psi-n\cos\psi)(\lambdao/L)\right]^2\right\}^{\frac{1}{2}}\,.
	\label{predict}
	\end{equation} 
	Since the thickness $\Ld$ is finite, shifts in the predictions of $\theta$ for
	specific values of $\lambdao$ and $\psi$ are possible for SPP waves. Also, shifts are  possible for both SPP waves and WGMs,
	because both canonical problems were formulated and solved with $\Lw=\La=0$. Finally, shifts can also be due to $\Lg\ne0$ \cite{Shuba2}. Therefore, 
	for all absorptance spectrums presented in this paper, we accepted predictions of $\theta$ from Eq.~(\ref{predict}) with $\pm1\deg$ tolerance.  However, let us note that not every possible GWM is strongly excited by planewave illumination.
	
	Finally, it is important to note that  depolarization can occur
	because the PCBR is doubly periodic. Accordingly, illumination
	by a linearly polarized plane wave for a specific value of $\psi$ can excite
	a GWM of a different polarization
	state propagating in a direction specified by the angle $\varphi$
	that may differ from $\psi$ \cite{9,Dutta-JOSAA}.

	\section{Numerical results and discussion}\label{nrd}
	
\begin{figure}[h]	
		\centering
		\includegraphics[scale=0.44]{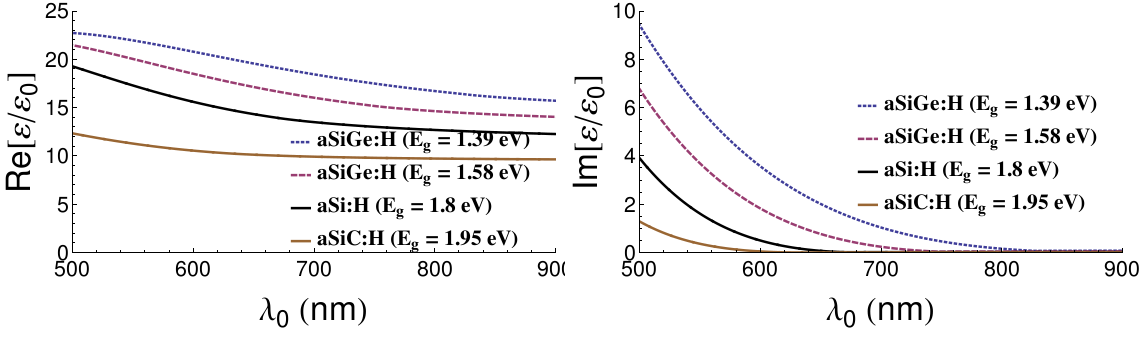}
		\caption{Spectrums of the relative permittivity $\eps/\epso$ of the different
			semiconductor alloys used in the triple-\Pin-junction tandem solar cell.
			\label{figure2}}
	\end{figure}

	All optical and geometric parameters were chosen  only to illustrate the relationships
		of the WGMs to total and useful absorptances, but still are representative of actual tandem
		solar cells \cite{FLML}.
	The compositions, bangaps, and  thicknesses of the nine
	hydrogenated a-Si alloys for the nine semiconductor layers are presented
	in Table~\ref{tab-1}.
	The  permittivity of each alloy was calculated as a function
	of $\lambdao$, using a model provided by Ferlauto et~al. \cite{6, 16}. The spectrums
	of all nine permittivities, normalized by $\epso$, are plotted in Fig.~\ref{figure2}. The 2D~PCBR was taken to be made of
	silver \cite{17}. The
	refractive index of AZO was taken as a function of $\lambdao$ from Gao \textit{et al.} \cite{AZO_data}. 
	
	\begin {table}[h]
	\caption { \label{tab-1} Compositions, bandgaps, and thicknesses of hydrogenated a-Si alloys
		used for the nine semiconductor layers in the triple-\Pin-junction tandem solar cell. 
	} 
	\begin{center}	
		\begin{tabular}{ p{1cm} p{3cm} p{2.5cm} p{2.4cm}  }
			\hline\hline
			Layer & Composition &Bandgap  (eV) & Thickness (nm) \\		
			\hline
			$\red{1p}$&  a-Si$_{1-u}$C$_u$:H &1.95&20 \\
			$\red{1i}$ & a-Si:H &1.8 & 200 \\
			$\red{1n}$ & a-Si:H &1.8 & 20 \\
			\hline
			$2p$& a-Si$_{1-u}$C$_u$:H&1.95&20 \\		
			$2i$&  a-Si$_{1-u}$Ge$_u$:H & 1.58 &200 \\
			$2n$ & a-Si:H &1.8 & 20 \\
			\hline
			$\red{3p}$ & a-Si:H &1.8 & 20 \\
			$\red{3i}$ & a-Si$_{1-u}$Ge$_u$:H &1.39 & 200\\
			$\red{3n}$ & a-Si:H &1.8 & 20 \\
			\hline\hline
		\end{tabular}
	\end{center}
	\end {table}

	The following dimensions were chosen: $\Lw=100$~nm, 
	$\La=60$~nm, $\Lg=80$~nm, $\Lm=30$~nm, $\Lx=\Ly=400$~nm, and $\zetax=\zetay=1$. 
	We used $M_t =N_t$ accordingly.
	Furthermore, we  used \red{$M_t\leq12$, which} ensured  the convergence of all non-zero 
	reflectances and   absorptances to within $\pm1\%$ for every 
	$\lambdao\in\left\{500,502,...,898,900\right\}$~nm. Here, convergence was
	defined to have occurred when there was a difference not exceeding 1\% in magnitude
	between the results for $M_t=N-1$ and $M_t=N$.
	Higher values of $M_t$  were found to be necessary for 
	higher $\lambdao$ as the chosen semiconductor alloys are then
	less absorbing and the effect of grating is more pronounced.

	\subsection{Prediction of GWM wavenumbers} \label{predGWM}
\begin{figure}[h]	
		
		\centering
		\includegraphics[scale=0.6]{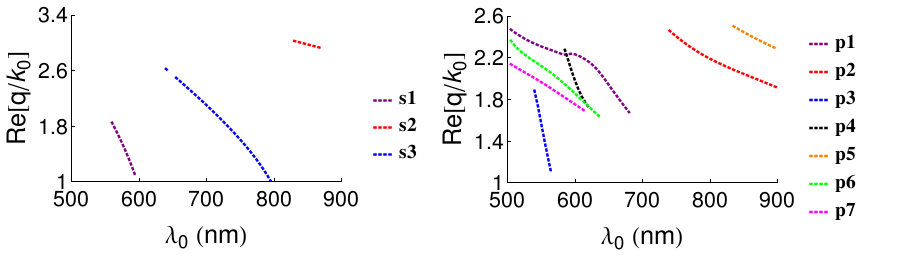}
		\caption{Real parts of the normalized wavenumbers $q/\ko$ of $s$- and $p$-polarized SPP waves obtained after solving the relevant canonical boundary-value problem. 
			\label{sppq}}
	\end{figure}

\begin{figure}[h]	
		\centering
		\includegraphics[scale=0.6]{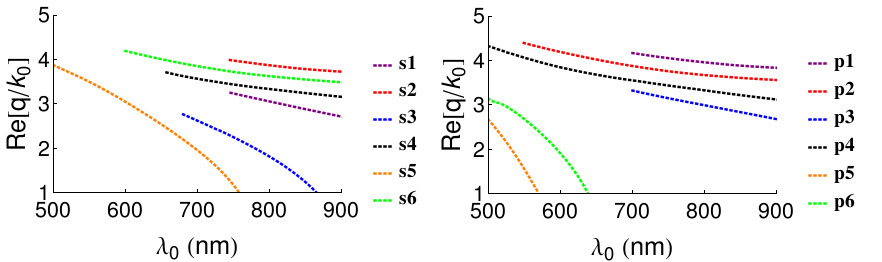}
		\caption{ Real parts of the normalized wavenumbers $q/\ko$ of $s$- and $p$-polarized WGMs obtained after solving the relevant canonical boundary-value problem.
			\label{wgmq}}
	\end{figure}

	The real parts of the normalized wavenumbers $ q/\ko$ of SPP waves are
	presented in Fig~\ref{sppq} as functions of $\lambdao\in\left\{500,501,...,899,900\right\}$~nm. These wavenumbers are organized into three branches (labeled $s1$--$s3$) for $s$-polarized SPP waves
	and seven branches (labeled $p1$--$p7$) for $p$-polarized SPP waves. The real parts of the normalized wavenumbers  $ q/\ko$ of the WGMs are presented in Fig~\ref{wgmq}.  These wavenumbers are arranged into six branches for both $s$- and $p$-polarized WGMs labeled $s1$--$s6$ and $p1$--$p6$, respectively.

	\subsection{\red{Absorptances} and Correlation with  Predictions}\label{abscorrel}
	Calculations of $\As$ and $\Ap$ as functions of $\lambdao\in[500,900]$~nm
	were made for the chosen triple-\Pin-junction tandem solar cell with a 2D~PCBR. In addition, we computed the spectrums of the
	useful  absorptances
	\begin{equation}
	\left.\begin{array}{l}
	\displaystyle{\Assc=\alphasc\Big\vert_{\bar{a}_{\rm p}=0}}
	\\[8pt]
	\displaystyle{\Apsc=\alphasc\Big\vert_{\bar{a}_{\rm s}=0}}
	\end{array}\right\}\,.
	\end{equation}

	The spectrums of $\As$,  $\Ap$, $\Assc$, and $\Apsc$ for
	$\lambdao\in[500,900]$~nm  for solar cells with and without corrugations ($\Lg=80$~nm and $\Lg=0$, respectively) were examined for several combinations of
	$\theta$ and $\psi$ \cite{Ahmad-SPIE2017}. For the sake of
	illustration, data are
	presented in Figs.~\ref{figure5}--\ref{figure8}
	only for the following two directions of incidence:
	\begin{itemize}
		\item[(1)] $\lec\psi=1\deg,\theta=1\deg\ric$, 
		and
		\item[(2)] $\lec\psi=45\deg,\theta=15\deg\ric$. 
	\end{itemize}
	The choices of $1\deg$ instead of $0\deg$ for the incidence angles help avoid  spurious results associated with the computation of distinct eigenvalues
	of $\*P(z)$ when the RCWA is implemented. Also shown in these figures
	are the spectrums of the useful absorptances $\bar{A}_{\rm s}^{{\rm sc}m}$
	and
	$\bar{A}_{\rm p}^{{\rm sc}m}$ in the $m$th \pin junction,
	$m \in\lec1,2,3\ric$, for incident $s$-
	and $p$-polarized plane waves, respectively.

	The excitation of a GWM is marked by an absorptance peak.
	Therefore, values
	of $\lambdao\in[500,900]$~nm  for which the
	solutions of the two canonical problems (with the assumption that
	$\Lw=\La=0$) predicted the excitation
	of SPP waves and WGMs for $\theta\in[0\deg,2\deg]\cup[14\deg,16\deg]$ are also
	identified in Figs.~\ref{figure5}--\ref{figure8}. Red arrows   indicate the excitation
	of SPP waves that matched with both total absorptances ($\As$ and $\Ap$) 
	\textit{and} useful
	absorptances ($\Assc$ and $\Apsc$); black arrows indicate
	WGMs that matched with both total absorptances \textit{and} useful absorptances; blue arrows indicate
	the excitation of SPP waves that  correlated with total absorptances but not with useful
	absorptances; and purple arrows indicate the  excitation of WGMs that 
	correlated with total absorptances but not with useful absorptances.


	\subsubsection{Case 1: $\lec\psi=1\deg,\theta=1\deg\ric$ }\label{case-1}
	
\begin{figure}[h]	
		\centering 
		\includegraphics[scale=0.44]{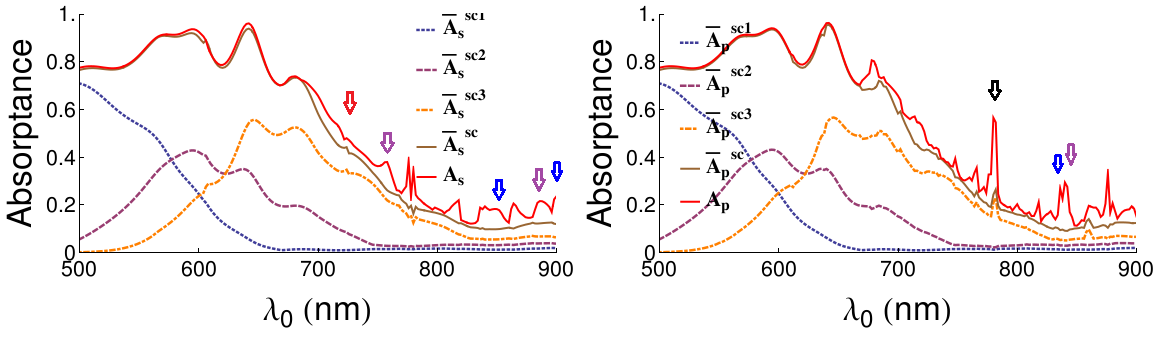}
		\caption{Spectrums of (left) $\As$, $\Assc$, $\Aassc$, $\Aaassc$, and $\Aaaassc$	
			and (right) $\Ap$, $\Apsc$, $\Aapsc$, $\Aaapsc$, and $\Aaaapsc$ 
			of the triple-\Pin-junction tandem solar cell, when 
			$\psi=1\deg$ and $\theta=1\deg$. 
			Red arrows indicate the excitation of SPP 
			waves that matched with both total absorptances ($\As$ and $\Ap$) and useful
			absorptances ($\Assc$ and $\Apsc$); black arrows indicate
			WGMs that matched with both total absorptances   and useful absorptances; 
			blue arrows indicate the excitation of SPP waves that  correlated with total absorptances 
			but not with useful absorptances;
			and purple arrows indicate the  excitation of WGMs that correlated with total 
			absorptances but not with useful absorptances.  
			\label{figure5}}
	\end{figure}

\begin{figure}[h]	
		
		\centering 
		\includegraphics[scale=0.44]{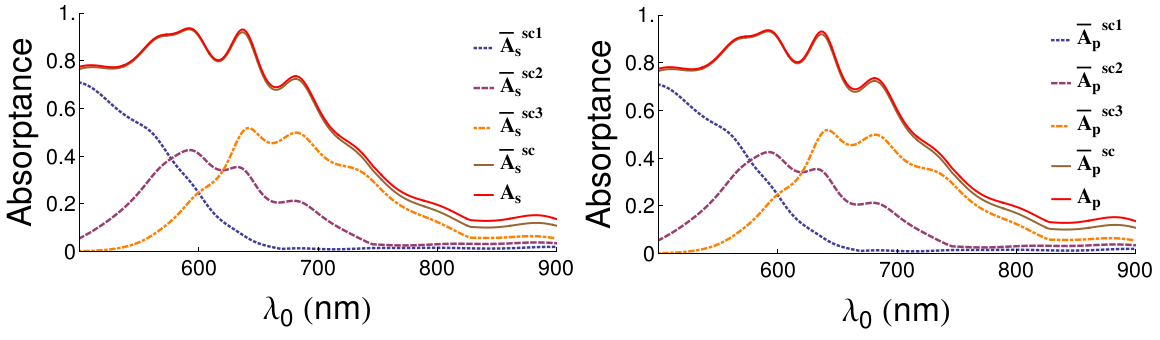}
		\caption{Same as Fig.~\ref{figure5} except that $\Lg=0$.
			\label{figure6}}
	\end{figure}

	Spectrums of  $\As$,  $\Ap$, $\Assc$ and $\Apsc$
	for $\lec\psi=1\deg,\theta=1\deg\ric$ calculated with $\Lg=80$~nm
	are presented in Fig.~\ref{figure5}. Also,   spectrums of the
	same quantities calculated with $\Lg=0$
	are presented in Fig.~\ref{figure6} for comparison.
	Tables~\ref{tab-2} and \ref{tab-3} contain values
	of $\lambdao\in[500,900]$~nm  for which the
	excitation of  either an SPP wave or a WGM
	as a Floquet harmonic of order $(m,n)$ is predicted.
	
	The $\Assc$-peak at $\lambdao\approx 728$~nm
	in Fig.~\ref{figure5} occurs close to the 
	wavelength $\lambdao\approx 730 $~nm predicted for the excitation of an
	$s$-polarized SPP wave
	as a Floquet harmonic of
	order $(1,0)$  at $\theta=0.856\deg$ in Table~\ref{tab-2}. This is the only
	SPP wave that   correlated with  peaks of both $\Assc$  and   $\As$. 
	
	The $\As$-peak  in Fig.~\ref{figure5} at 
	\begin{itemize} 		
		\item
		$\lambdao\approx 845$~nm is due to
		the excitation of an $s$-polarized SPP wave as a Floquet harmonic of
		order $(1,1)$ predicted at $\theta=0.088\deg$ in Table~\ref{tab-2},
		\item
		$\lambdao\approx 897$~nm matches well with the excitation of a $p$-polarized
		SPP wave as a Floquet harmonic of
		order $(1,0)$ predicted at $\theta=0.847\deg$ in Table~\ref{tab-2},
		\item 
		$\lambdao\approx 754$~nm is related with
		the excitation of a $p$-polarized WGM as a Floquet harmonic of
		order $(-2,0)$ predicted at $\theta=1.145\deg$ in Table~\ref{tab-3},
		and 
		\item 
		$\lambdao\approx 892$~nm is represents
		the excitation of a $p$-polarized WGM as a Floquet harmonic of
		order $(-1,1)$ predicted at $\theta=1.198\deg$ in Table~\ref{tab-3}, 
	\end{itemize}
	Excitation of these GWMs  correlated only with the total absorptance $\As$ but not with the  useful absorptance $\Assc$,
	which indicates that not every $\Assc$-peak can be matched to an $\As$-peak
	that is correlated with the excitation of a GWM {\cite{9,FLpra2011}}. Accordingly, 
	useful absorptance, not the overall absorptance, needs to be studied 
	for solar cells. Contributions to the overall absorptance are made both by 
	the semiconductor layers and the metallic PCBR, but the  contribution of the latter
	is useless for harvesting solar energy.


	The $\Apsc$-peak in Fig.~\ref{figure5} at 
	$\lambdao\approx 786$~nm is due to
	the excitation of an $s$-polarized WGM  as a Floquet harmonic of
	order $(-1,0)$ predicted at $\theta=0.948\deg$ in Table~\ref{tab-3}.
	The $\Ap$-peak   at 
	\begin{itemize} 		
		\item
		$\lambdao\approx 834$~nm is due to
		the excitation of a $p$-polarized WGM as a Floquet harmonic of
		order $(-1,1)$ predicted at $\theta=0.838\deg$ in Table~\ref{tab-3}, and
		\item
		$\lambdao\approx 845$~nm matches well with the excitation of a $p$-polarized
		SPP wave as a Floquet harmonic of
		order $(1,1)$ predicted at $\theta=0.088\deg$ in Table~\ref{tab-2},
	\end{itemize}

	\begin {table}[h]
	\caption {\label{tab-2} Values
		of $\lambdao\in[500,900]$~nm (calculated at $1$-nm intervals) for which the 
		excitation of  an SPP wave as a Floquet harmonic of order $(m,n)$ is predicted
		for $\theta\in[0\deg,2\deg]$  and $\psi=1\deg$, for the tandem solar cell with a
		2D~PCBR. The SPP waves strongly excited in Fig.~\ref{figure5} are highlighted in bold.}
	
	\begin{center}
		\begin{tabular}{p{1.6cm} p{1.2cm} p{1.6cm} p{1cm} p{1.25cm}}
			\hline
			\hline
			Pol. State  & $\lambdao$~(nm) & ${\rm Re}\{q/\ko\}$ &$\theta\deg$ 
			& $(m, n)$ \\
			\hline	
			$\textbf{s}$&$\mathbf{730}$&	$\mathbf{1.816}$&$\mathbf{	0.856}$&$\mathbf{(1, 0)}$\\
			$\textbf{s}$&$\mathbf{845}$&	$\mathbf{2.988}$&	$\mathbf{0.088}$&$\mathbf{(1, 1)}$\\
			$p$&570&	2.027&	0.991&$(1, 1)$\\
			$p$&640	&2.029	&1.024	&$(\pm1, 0)$\\
			$p$&680&1.678&1.228&$(1, 0)$\\
			$\textbf{p}$&$\mathbf{897}$&$\mathbf{2.988}$&$\mathbf{0.847}$&$\mathbf{(1, 0)}$\\	
			\hline
			\hline
		\end{tabular}
	\end{center}
	\end {table}

	\begin {table}[h]
	\caption {\label{tab-3} Same as Table~\ref{tab-2}, except that the relevant excitations 
		of WGMs are indicated. The WGMs strongly excited in Fig.~\ref{figure5} 
		are highlighted in bold.} 
	\begin{center}
		\begin{tabular}{p{1.6cm} p{1.2cm} p{1.6cm} p{1cm} p{1.2cm}}
			\hline
			\hline
			Pol. State  & $\lambdao$~(nm) & ${\rm Re}\{q/\ko\}$ &$\theta\deg$ 
			& $(m, n)$ \\
			\hline	
			$s$&663&	3.685&	1.334&$(-2, 1)$\\	
			$s$&711&	3.536&	1.076&$(-2, 0)$\\
			$\textbf{s}$&$\mathbf{786}$&	$\mathbf{1.948}$&	$\mathbf{0.948}$&$\mathbf{(-1, 0)}$\\	
			$\textbf{s}$&$\mathbf{834}$&$\mathbf{2.938}$&$\mathbf{0.838}$&$\mathbf{(-1, 1)}$\\		
			$s$&898		&3.163&0.902&$(-1, 1)$\\
			$\textbf{p}$&$\mathbf{754}$&$\mathbf{3.750}$&	$\mathbf{1.145}$&$\mathbf{(-2, 0)}$\\
			$p$&797&3.963&1.206&$(-2, 0)$\\
			$p$&827 &2.910 &1.121  &$(-1, 1)$ \\
			$\textbf{p}$&$\mathbf{892}$&$\mathbf{3.138}$ &$\mathbf{1.198}$&$\mathbf{(-1, 1)}$\\
			\hline
			\hline
		\end{tabular}
	\end{center}
	\end {table}

	On comparing Figs.~\ref{figure5} and \ref{figure6}, we note that the GWMs are
	excited at $\lambdao>700$~nm.  Also, the total absorptance for $\Lg=80$~nm
	exceeds that for $\Lg=0$ in the same spectral regime. 
	This increase is largely due to the increases in $\Aaaassc$ and $\Aaaapsc$, i.e., in the \pin junction closest
	to the PCBR. 
	Furthermore,  increases in both  total and useful absorptances   
	for $\lambdao\in[634,680]$~nm, regardless of the polarization state of the incident light, were observed with the use of the  PCBR  rather than a planar backreflector.
	In addition,  depolarization  due to the two-dimensional periodicity of the PCBR is evident from the excitation of WGMs that are not of the same polarization state as the incident light.



	\subsubsection{Case 2: $\lec\psi=45\deg,\theta=15\deg\ric$}\label{case-2}

\begin{figure}[h]	
		\centering 
		\includegraphics[scale=.44]{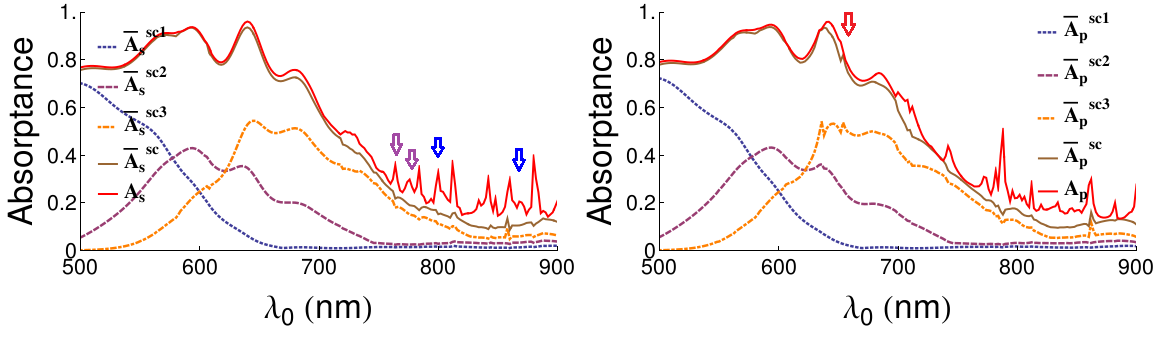}
		\caption{Same as Fig.~\ref{figure5}, except that $\theta=15\deg$ and $\psi=45\deg$ .
			\label{figure7}}
	\end{figure}

\begin{figure}[h]	
		\centering 
		\includegraphics[scale=.44]{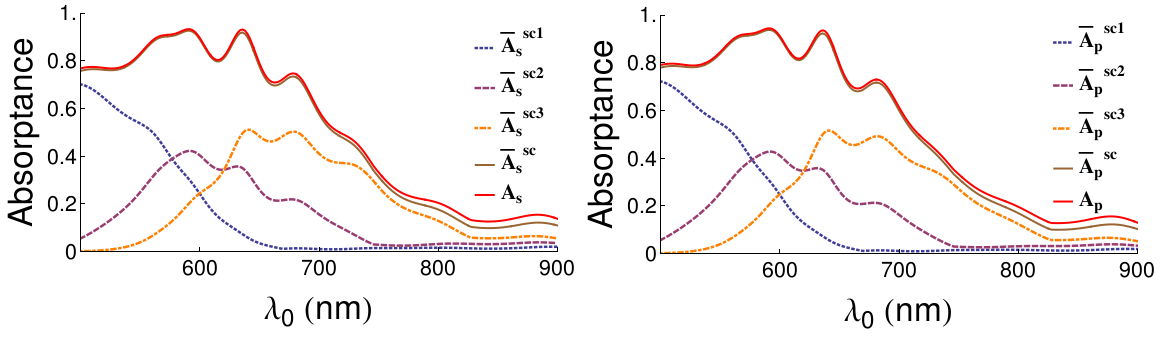}
		\caption{Same as Fig.~\ref{figure7}, except that $\Lg=0$.
			\label{figure8}}
	\end{figure}

	Calculated spectrums of $\As$, $\Ap$, $\Assc$, and $\Apsc$
	for $\lec\psi=45\deg,\theta=15\deg\ric$ with $\Lg=80$~nm
	are presented in Fig~\ref{figure7}. Also, the  spectrums $\As$, $\Ap$, $\Assc$, and $\Apsc$ calculated with $\Lg=0$ 
	for the same incident direction are presented 
	in Fig.~\ref{figure8}. Tables~\ref{tab-4}
	and \ref{tab-5} contain values
	of $\lambdao\in[500,900]$~nm for which the
	excitation
	of  either an SPP wave or a WGM
	as a Floquet harmonic of order $(m,n)$
	is predicted from  analysis of Figs.~\ref{sppq} and \ref{wgmq}..
	
	No $\Assc$-peak could be correlated with the excitation of a GWM.  The 
	$\As$-peak at
	\begin{itemize}
		\item	
		$\lambdao\approx 800$~nm is related to
		the excitation of a $p$-polarized SPP wave as a Floquet harmonic of
		order either $(1,0)$ or $(0,1)$ predicted at $\theta=14.952\deg$ in Table~\ref{tab-4},
		\item	
		$\lambdao\approx 870$~nm is due to
		the excitation of a $p$-polarized SPP wave as a Floquet harmonic of
		order either $(-1,0)$ or $(0,-1)$ predicted at $\theta=15.880\deg$ in Table~\ref{tab-4},
		\item
		$\lambdao\approx 764$~nm is associated with
		the excitation of a $p$-polarized WGM  as a Floquet harmonic of
		order $(-1,-2)$ predicted at $\theta=15.141\deg$ in Table~\ref{tab-5}.
		and
		\item
		$\lambdao\approx 778$~nm arises due to
		the excitation of a $p$-polarized WGM  as a Floquet harmonic of
		order either $(-2,0)$ or $(0,-2)$ predicted at $\theta=15.411\deg$ in Table~\ref{tab-5}.
	\end{itemize}
	The $\Apsc$-peak at	$\lambdao\approx 647$~nm is related with
	the excitation of an $s$-polarized SPP wave as a Floquet harmonic of
	order $(1,1)$ at $\theta=15.529\deg$ in Table~\ref{tab-4}. 
	No other \red{$\Apsc$-} or $\Ap$-peak \red{was found to be correlated} with GWM excitation. 
	These results underscore the fact that useful absorptance is not necessarily enhanced by the excitation of a GWM. However,
	there are \red{useful- and total-absorptance} peaks
	which could not predicted 
	by the canonical boundary-value problems.

	\begin {table}[h]
	\caption {\label{tab-4} Values
		of $\lambdao\in[500,900]$~nm (calculated at $1$-nm intervals) for which the excitation of  
		an SPP wave as a Floquet harmonic of order $(m,n)$ is predicted for $\theta\in[0\deg,2\deg]$  
		and $\psi=45\deg$, for the tandem solar cell backed by a 2D PCBR. The SPP waves strongly 
		excited in Fig.~\ref{figure7}
		are highlighted in bold.} 
	\begin{center}
		\begin{tabular}{p{1.6cm} p{1.2cm} p{1.6cm} p{1cm} p{2.2cm}}
			\hline
			\hline
			Pol. State  & $\lambdao$~(nm) & ${\rm Re}\{q/\ko\}$ &$\theta\deg$ 
			& $(m,n)$ \\
			\hline	
			$\textbf{s}$&$\mathbf{647}$&$\mathbf{2.581}$&$\mathbf{15.529}$	&$\mathbf{(1, 1)}$\\	
			$s$&	690&	2.193	&14.231	&$(-1, -1)$\\		
			$p$&	565&	2.272	&15.955	&$(1, 1)$\\
			$p$&620&	1.734	&14.365	&$(1, 0), (0, 1)$\\
			$p$&	750&	2.404	&14.315	&$(-1, -1)$\\
			$\textbf{p}$&$\mathbf{800}$&	$\mathbf{2.190}$	&$\mathbf{14.952}$	&$\mathbf{(1, 0)}, \mathbf{(0, 1)}$\\
			$\textbf{p}$&$\mathbf{870}$&	$\mathbf{2.376}	$&$\mathbf{15.880}$	&$\mathbf{(-1, 0)}, \mathbf{(0, -1)}$\\
			\hline
			\hline
		\end{tabular}
	\end{center}
	\end {table}

	\begin {table}[h]
	\caption {\label{tab-5} Same as Table~\ref{tab-4}, except that the relevant excitations 
		of WGMs are indicated. The WGMs strongly excited in Fig.~\ref{figure7}
		are highlighted in bold.} 
	\begin{center}
		\begin{tabular}{p{1.6cm} p{1.2cm} p{1.6cm} p{1cm} p{2.2cm}}
			\hline
			\hline
			Pol. State  & $\lambdao$~(nm) & ${\rm Re}\{q/\ko\}$ &$\theta\deg$ 
			& $(m, n)$ \\
			\hline	
			$s$&	732&	3.484	&15.004	&$(-2, 0), (0, -2)$\\
			$s$&	807&	3.858	&14.931	&$(-2, 0), (0, -2)$\\
			$\textbf{p}$&$\mathbf{764}$&$\mathbf{4.023}$	&$\mathbf{15.141}$	&$\mathbf{(-1, -2)}$\\
			$\textbf{p}$&$\mathbf{778}$&$\mathbf{3.706}$    &$\mathbf{15.411}$    &$\mathbf{(-2, 0)}, \mathbf{(0, -2)}$\\
			$p$&	664&	3.641	&14.287	&$(-2, 1)$\\
			$p$&	732&	3.477	&15.355	&$(-2, 0), (0, -2)$\\
			$p$&	821&	3.925	&15.098	&$(-2, 0), (0, -2)$\\
			\hline
			\hline
		\end{tabular}
	\end{center}
	\end {table}

	On comparing Figs.~\ref{figure7} and \ref{figure8},  increases in both  total and useful absorptances   
	for $\lambdao\in[640,670]$~nm, regardless of the polarization state of the incident light, become evident with the use of the  PCBR  rather than a planar backreflector. In comparison to normal illumination (Sec.~\ref{case-1}) for which GWMs were excited only for $\lambdao>700$~nm, an SPP wave is excited at $\lambdao=647$~nm for oblique illumination. This is in accord with 
	the  blueshifts of SPP waves expected for oblique illumination \cite{6,Dutta-JOSAA} as well as with the angular trends in Figs.~\ref{sppq} and \ref{wgmq}.
	
	Apart from the SPP wave excited at $\lambdao=647$~nm,
	all other GWMs are
	excited at $\lambdao>700$~nm. The polarization state of an excited GWM may not be the same as that of the incident light, because the 2D PCBR is a depolarizing agent.
	Finally, the total absorptance increases in the same spectral regime with the use of the PCBR in comparison to a planar backreflector,  for either polarization
	state of the incident light, largely due to the increases in $\Aaaassc$ and $\Aaaapsc$, i.e., in the \pin junction closest
	to the PCBR.

	\section{Concluding remarks}\label{conc}

	The effect of a 2D~PCBR on the  absorptance of light
	in a triple-\Pin-junction thin-film solar cell was studied using the RCWA. 	
	Total absorptances and useful absorptances
	for incident $s$- and $\ppol$ light were computed against the free-space wavelength
	for two different incidence directions. 
	Calculations were also made of the useful absorptance 
	in each of the three \Pin~junctions. Furthermore, two  canonical boundary-value problems
	were solved for the prediction of GWMs. The predicted GWMs were correlated 
	with the peaks of the total and useful absorptances for both linear polarization states. 
	
	Numerical studies led to the following conclusions:
	\begin{itemize}
		
		\item
		Regardless of the illumination direction and the polarization state of the
		incident light,  increases in useful and total absorptances for $\lambdao<700$~nm
		arise from the replacement of a planar backreflector by a 2D PCBR.
		
		\item
		The triple-\Pin-junction tandem solar cell made of a-Si alloys is 
		highly absorbing for $\lambdao<700$~nm, so that the
		excitation of SPP waves in this regime is unnoticeable.
		
		\item
		Both SPP waves and WGMs are excited for $\lambdao>700$~nm for both normal and oblique illumination. An
		SPP wave excited at $\lambdao=647$~nm for oblique illumination is in accord with blueshifting of SPP waves 
		with increasing obliqueness of illumination.

		\item 
		Some of the excited GWMs directly contribute to the increase in useful 
		absorptance of the solar cell backed by a 2D PCBR.
		This increase is largely due to enhanced absorptance
		in the \pin junction closest
		to the 2D PCBR.
		
		\item
		Depolarization  due to the two-dimensional periodicity of the PCBR is evident 
		from the excitation of GWMs that are not of the same polarization state as the incident light.
		
		\item
		Excitation of certain GWMs  could be correlated  with the total absorptance but not with  the useful absorptance. 
		
	\end{itemize}
	When devising light-trapping strategies, the
	useful, but not the total absorptance needs to be focused on.  While reduction of reflectance is a worthwhile objective, meeting it will not necessarily boost the useful absorptance.
	We conclude with the recommendation that   material and geometric parameters need to be optimized for efficiency enhancement.

	\noindent {\bf Note.} This paper is substantially based on a paper titled, ``On optical-absorption peaks in a
	nonhomogeneous dielectric material over a two-dimensional metallic
	surface-relief grating," presented at the SPIE Optics and Photonics conference Nanostructured Thin Films X, held 
	August 5--11, 2017 in San Diego, California, United States. \\
	
	\noindent \textbf{Acknowledgments.} F. Ahmad thanks the Graduate School and the College of Engineering,  Pennsylvania State
	University, for a University Graduate Fellowship during the first year of his doctoral
	studies. A. Lakhtakia thanks the Charles Godfrey Binder Endowment at the Pennsylvania
	State University for ongoing support of his research. The research of  F. Ahmed and A. Lakhtakia is partially supported by  US National Science
	Foundation (NSF)  under grant number DMS-1619901. The research of T.H. Anderson, B.J. Civiletti, 
	and P.B.  Monk is partially supported by  the US National Science Foundation (NSF) under 
	grant number DMS-1619904.


\end{document}